# Linear Dielectric Thermodynamics:

# A New Universal Law for Optical, Dielectric Constants

by

S. J. Burns

Materials Science Program

Department of Mechanical Engineering

University of Rochester

Rochester, NY 14627 USA

## *Abstract*

Linear dielectric thermodynamics are formally developed to explore the isothermal and adiabatic temperature - pressure dependence of dielectric constants. The refractive index of optical materials is widely measured in the literature: it is both temperature and pressure dependent. The argument to establish the dielectric constant's isentropic temperature dependence is a thermodynamic one and is thus applicable to all physical models that describe electron clouds and electronic resonances within materials. The isentropic slope of the displacement field versus the electric field at all temperatures is described by an adiabatic dielectric constant in an energy-per-unit mass system. This slope is shown through the electronic part of the entropy to be unstable at high temperatures due to the change in the curvature of the temperature dependence of the dielectric constant. The electronic entropy contribution for optical, thermo-electro materials has negative heat capacities which are unacceptable. The dielectric constant's temperature and pressure dependence is predicted to be only dependent on the specific volume so isentropes are always positive. A new universal form for the dielectric constant follows from this hypothesis: the dielectric constant is proportional to the square root of the specific volume.

*Introduction*

Dielectric materials are physically modeled through the electric susceptibility with the atomic polarization density determined by: electronic polarization produced by opposite displacements of electronic clouds and positive nuclei; ionic polarizability caused by opposite displacements of positive and negative ions; orientational contributions of complex ions or molecules from permanent dipoles in the material. Since the electron's and molecule's positions are temperature dependent, the electronic processes that create permittivity depend on temperature. Displacements in crystals are also pressure dependent so the dielectric constant is a function of both temperature and pressure. The dielectric constant is not constant but is well known to contain both static and frequency dependent parts as well as the temperature and pressure dependence. The frequency dependence of the dielectric constant is understood for the electronic, dipolar and ionic parts as is discussed in condensed matter and electromagnetic texts [1-4].

The accepted physical modelling of molecular electric permittivity was discovered independently by two scientists with similar names: Lorentz [5] and Lorenz [6], and is accordingly called the Lorentz-Lorenz expression. Each active charge in the Lorentz-Lorenz description is a center for providing local polarization. Dispersion is the variation of the refractive index with frequency and it comes from the charge cloud resonances in dielectric media; it is well known but complicated by electronic motions, resonances and damping within the material. The resonances and damping especially at optical frequencies are generally described by Sellmeier-based [7-9] equations. Glass manufacture's use a slightly expanded form of this expression for the damped resonances of the refractive index to describe their materials, as is well explained in [10]. The theoretical physical models of dielectric solids (as opposed to gases) have very limited discussions of the entropy and temperature changes within dielectric media. At optical frequencies and wavelengths the media are unlikely to be excited in an isothermal, thermodynamic state.

Optical materials for the next generation of optical devices [11-14] will rely heavily on the physical properties of structures that are yet to be designed using materials at their extreme limits to achieve functionality. The optical properties often must vary within the media in controlled structures. Temperature and dispersion characteristics of the dielectric media are very important to systematic design and construction of cloaking devices [11-13], freeform optics [14] and the yet to be made, next generation of super-high intensity laser glasses. Again, at optical frequencies and wavelengths these media are yet to be described by a thermodynamic state.

A formal thermodynamic energy balance per unit mass is used here to find the entropy in isothermal, electro-polarized materials. Changes in the dielectric constant with temperature in this thermodynamic system are shown to determine the electro-entropy. The calculated entropy is found via the thermodynamic Maxwell's relation which determines the contribution to the heat capacity due to the applied electric field. The isotherms which are excellent descriptions of linear dielectrics are well known in electro-dynamic materials. All linear dielectrics display a point where isotherms cross which in turn has consequences for the second law of thermodynamics. The adiabatic curves, if thought to follow the isotherms through the crossing-point, will have a second law violation by the Kelvin-Planck cycle; Carathéodory's Principle is also violated. It is shown in the glasses studied that the dielectric constant's temperature dependence poses a problem with the heat capacity. It is proposed that when the dielectric constant is in a "universal" form then the material is stable, and no second law violations are perpetuated. Empirical evidence presented here supports the adiabatic dielectric constant found from the thermodynamic analysis developed below.

*Thermodynamic Energy Balance in an Electric Field*

Figure 1 shows a schematic of the electric displacement field, $D$ and electric field, $E$ at temperatures, $T_1$, $T_2$, $T_3$ and $T_4$. The isothermal lines are related by the equation:

$$D = \varepsilon E, \qquad (1)$$

where $\varepsilon$ is the electric permittivity. In general, it is a tensor that connects the displacement and electric field vectors. In what follows below, $E$ and $D$ are scalars and $\varepsilon$ is considered as an isotropic scalar that has explicit $T$ and $p$ dependence; $T$ is the absolute temperature and $p$ is the pressure.

$$\varepsilon = \varepsilon(T, p) \qquad (2)$$

Equation (1) plus (2) is an equation of state for the independent variables $E$, $T$ and $p$.

Well-known expressions for optical materials relate the dielectric constant $\kappa$, the permittivity of free space, $\varepsilon_0$ and the refractive index, $n$ as

$$\kappa \equiv \frac{\varepsilon}{\varepsilon_0}; \quad n = \sqrt{\kappa}. \qquad (3)$$

A full formal thermodynamic system assuming that equations (1) and (2) are valid is developed below. The analysis determines the adiabatic equation of state consistent with equations (1) and (2). The analysis presented is a thermodynamic one so it organizes all the physical properties without prior assumptions and allows for transformation of variables to other descriptions through Jacobian algebra. The thermodynamic analysis is systematic so it keeps assumptions and physical properties in full perspective.

The first law of thermodynamics in incremental form for dielectric media is an energy balance per unit mass given by:

$$du = Tds - pdv + vEdD \qquad (4)$$

$u$ is the internal energy per unit mass, $s$ is the entropy per unit mass and $v$ is the specific volume per unit mass. $Tds$ is the incremental heat added to the system, $pdv$ is the incremental mechanical work done by the system on the surroundings while $vEdD$ is the incremental electrical work done on the system. The free energy per unit mass, g is a Gibbs-like function:

$$g \equiv u - Ts + pv - E\underline{D}. \quad (5)$$

A new term in equation (5) is the displacement-volume, $\underline{D}$. It is defined only in incremental form.

$$d\underline{D} \equiv vdD. \quad (6)$$

So from equation (4) through (6) we find

$$dg = -sdT + vdp - \underline{D}dE. \quad (7)$$

Note that equation (7) is in units of energy per unit mass. Most (maybe all) energy balances in electrical systems are in energy per unit volume [2, 9, 15-20] but the use of $\underline{D}$ in equation (5) allows for the energy per unit mass in isothermal, isobaric, isoelectric systems to be determined. $\underline{D}$ does not seem to have been used in the literature except briefly in [21].

The general non-linear description of the thermodynamic system has 6 physical properties. They are found from derivatives of the dependent variables with respect to the independent variables as is now outlined in equation (8) through (13).

The isobaric, isoelectric, heat capacity per unit mass, $C_{p,E}$, is defined in equation (8). In the general case, all the properties below are dependent on the three independent variables in the system: $T$, $p$ and $E$.

$$T \left.\frac{\partial s}{\partial T}\right|_{p,E} \equiv C_{p,E}. \quad (8)$$

The isobaric, isoelectric, volumetric thermal expansion coefficient is:

$$\frac{1}{v}\left.\frac{\partial v}{\partial T}\right|_{p,E} \equiv \alpha. \quad (9)$$

The isothermal, isoelectric, compressibility is:

$$-\frac{1}{v}\left.\frac{\partial v}{\partial p}\right|_{T,E} \equiv \beta. \quad (10)$$

The isothermal, isobaric, electric permittivity is:

$$\left.\frac{\partial D}{\partial E}\right|_{T,p} \equiv \varepsilon. \tag{11}$$

Note, equation (1) is for linear dielectrics while equation (11) applies to non-linear materials.

The isobaric, isoelectric, temperature dependence of the displacement field is:

$$\left.\frac{\partial D}{\partial T}\right|_{p,E} \equiv c_1. \tag{12}$$

And the isothermal, isoelectric, pressure dependence of the displacement field is:

$$\left.\frac{\partial D}{\partial p}\right|_{T,E} \equiv c_2. \tag{13}$$

Table I is a list of definitions from equations (8) through (13) and agrees with conventional thermo-electric systems. The expressions in Table I are in a Jacobian format. The table is symmetric about the diagonal because Maxwell's relations from the mixed derivatives in either order of the Gibbs-like function are equal. For example,

$$\frac{\partial^2 g}{\partial p \partial T} = -\left.\frac{\partial s}{\partial p}\right|_{T,E} = \left.\frac{\partial v}{\partial T}\right|_{p,E} = \frac{\partial^2 g}{\partial T \partial p}. \tag{14}$$

$c_1$ for linear dielectrics is found using equation (1) and (12) above:

$$c_1 = E \left.\frac{\partial \varepsilon}{\partial T}\right|_p. \tag{15}$$

$c_2$ is found for linear dielectrics using equation (1) and (13):

$$c_2 = E \left.\frac{\partial \varepsilon}{\partial p}\right|_T. \tag{16}$$

Table I is useful for finding derivatives that are not listed by changing variables with the assistance of Jacobian algebra. For example, consider the ratio of the heat to the electrical work for an isothermal, isobaric, dielectric material. This ratio is an indication of the heat exchanged with the media's surroundings in linear dielectrics during isothermal electrification of the media.

$$\left.\frac{T}{E}\frac{\partial s}{\partial \underline{D}}\right|_{T,p} = \frac{T}{E}\frac{J(s,T,p)}{J(\underline{D},T,p)} = \frac{T}{E}\frac{\begin{vmatrix} C_{p,E}/T & -v\alpha & vc_1 \\ 1 & 0 & 0 \\ 0 & 1 & 0 \end{vmatrix}}{\begin{vmatrix} vc_1 & vc_2 & v\varepsilon \\ 1 & 0 & 0 \\ 0 & 1 & 0 \end{vmatrix}} = \frac{Tc_1}{E\varepsilon} = \left.\frac{\partial(\ell n\varepsilon)}{\partial(\ell nT)}\right|_p. \qquad (17)$$

The Jacobian entries are read into the determinant from the rows in Table I. These entries are shown in the middle equation and this expression is evaluated using the Jacobian determinant. $c_1$ is eliminated using equation (15) and does not appear in the term on the extreme right. The absolute incremental ratio of the heat to the mechanical work for an ideal gas is 1. Linear dielectrics have a material specific ratio determined by the slope on a log-log plot, restricted to an isobar of the dielectric constant *versus* the temperature. The temperature dependence of the permittivity for a linear dielectric is evaluated through the dielectric constant's temperature dependence which is reported through the temperature dependence of the index of refraction seen in Table II. This ratio was evaluated using Schott's Data Catalog [22]. For one example, BK-7 has a value of $1.3 \times 10^{-3}$. Thermal effects in optical dielectrics are quite small. The heat energy from electrification in optical dielectric glasses is very poorly described by ideal gas concepts.

A second example is the temperature changes in a linear dielectric with application of an electric field with isentropic conditions:

$$\left.\frac{\partial T}{\partial E}\right|_{s,p} = \frac{J(T,s,p)}{J(E,s,p)} = -vE\frac{T}{C_{p,E}}\left.\frac{\partial \varepsilon}{\partial T}\right|_p \qquad (18)$$

so

$$\int_{T_0}^{T_1} C_{p,E}\frac{dT}{T} = -\int_0^E v\left.\frac{\partial \varepsilon}{\partial T}\right|_p EdE \qquad (19)$$

or

$$-\frac{v}{2}\left.\frac{\partial \varepsilon}{\partial T}\right|_p E^2 \approx C_{p,E}\ell n(\frac{T_1}{T_0}) = C_{p,E}\ell n(1+\frac{\Delta T}{T_0}) \approx C_{p,E}\frac{\Delta T}{T_0}. \qquad (20)$$

Use has been made of $\Delta T = T_1 - T_0$.

$$\Delta T = -\frac{v\varepsilon_0}{C_{p,E}} \frac{E^2}{2} T_0 \left.\frac{\partial k}{\partial T}\right|_p. \qquad (21)$$

The thermal change, $\Delta T$, in a linear dielectric is of the opposite sign from the dielectric constants' temperature dependence multiplied by the absolute temperature and proportional to the vacuum energy of the electric field divided by the heat capacity. The separation of equation (18) and (19) for integration which results in (20) is approximate and shows that some small part the electric field energy is modified by a term that is the temperature dependence of the dielectric constant times the absolute temperature. This energy changes the temperature of the material through the heat capacity times the change in temperature. $T$ changes with the entropy held constant, so there is no heat. See Table II for data on selected glasses; the electrical energy results in an adiabatic change in $\Delta T$; this energy is reduced to about 1 % or less by the last two terms in (21). The assumption used was that $\Delta T$ over $T_0$ is small and all integrations in (19) are over small ranges. The expressions developed above have no intrinsic dissipation as the system is reversible.

The optical medium between capacitor plates changes $T$ upon rapid electrification as seen above. It will increase or decrease the temperature depending on the temperature dependence of the permittivity or the $T$ dependence of the dielectric constant as given in equation (21). The capacitor's temperature will always equilibrate by moving heat from the surroundings until thermal equilibrium is achieved and the process can be described as isothermal with a uniform temperature. If however the electrification mechanism is say in an alternating field at optical frequencies where wavelengths in the media covers 1,000s of atoms and the times are computed in optical cycles, then there is not sufficient time to equilibrate the temperature in the system and the processes are considered isentropic; if it is insulated from the surroundings it is adiabatic. Figure 2 shows one cycle of an optical wave with $\Delta T$ changes, following expression (21).

The electro-magnetic wavelengths at X-ray frequencies are on atomic scales with heat lengths that are also on atomic scales. Thermodynamic processes at X-ray frequency ranges are necessarily isothermal. Dielectric constants at X-ray wavelengths are well known to be slightly less than 1.

*Electric Field Contribution to the Entropy*

The electric and displacement fields in equation (1) imply the isothermal, electrical permittivity seen in figure 1. The analysis used here has as an objective to find the isentropic curve as developed below. The entropy contribution in Table I is evaluated from the terms on the $s$ row of Table I.

$$ds = ds_1 + ds_2 + ds_3. \qquad (22)$$

The electric field's entropy is $s_3$, using

$$\left.\frac{\partial \mathbf{s}_3}{\partial E}\right|_{T,p} = vc_1 = vE \left.\frac{\partial \varepsilon}{\partial T}\right|_p. \qquad (23)$$

The expression in equation (23) on the right is for linear dielectrics; this equation is written in integral form below using the differential expression.

$$\int d\mathbf{s}_3 = \int v \left.\frac{\partial \varepsilon}{\partial T}\right|_p E dE. \qquad (24)$$

The integration of equation (24) takes place with $T$ and $p$ constant as specified on the left in equation (23). $v$ and $\varepsilon$ are functions of $T$ and $p$ so they can be taken out from under the integral sign. Thus,

$$\mathbf{s}_3 = v \frac{E^2}{2} \left.\frac{\partial \varepsilon}{\partial T}\right|_p. \qquad (25)$$

$\mathbf{s}_3$ is the electric field's entropy contribution to the system. The factor of 2 and the square in the electric field arises because it is a linear dielectric system; $v$ in the expression comes about because the system is based on energy per-unit-mass. Equation (25) is quadratic in $E$ and connects entropy and $T$ on an isobar. The Maxwell's stress tensor [18-20, 23] connects stresses and strains on an isotherm with quadratic $E$ fields in linear dielectric systems. Reference [23] has the Maxwell stress tensor and the entropy expression: equation (25) on their page 47 is the entropy in an energy-per-unit-volume system and the Maxwell's stress tensor is on page 29.

Equations (1) and (25) are now used to describe the constant $\mathbf{s}_3$ curve associated with figure 1. Assume that $\varepsilon = \varepsilon(T, p)$ from equation (2). Furthermore, $\varepsilon$ or $\kappa$ is assumed to be a known function of $T$ when measured on an isobar. Thus, the $T$ dependence of $\kappa$ as obtained below is from measured values. The method used here is to choose a value of $\mathbf{s}_3$ and $T$. Appendix A includes a graph of $v$'s $T$ dependence and how it is related to measured expansivity. The emphasis in this and the next section is to concentrate on the $T$ dependence found on isentropic curves. These curves are constructed parametrically in $T$; a value of $T$ is selected which allows $v$ and $d\kappa/dT$ to be determined. $E$ is now obtained from equation (25) after solving for $E$. See equation (26) below. The value of $T$ also determines $\varepsilon$ in equations (1) and (2) allowing $D$ to be found from the known values of $E$, $\varepsilon_0$, $v$, $\kappa$ and $dk/dT$. $D$ is given in equation (27) below. $D$ and $E$ values are thus established at particular $T$ and $\mathbf{s}_3$ values. Neighboring $D$ and $E$ values are obtained from $T + \Delta T$ with the same value of $\mathbf{s}_3$. The entire curve of $\mathbf{s}_3$ is constructed following this procedure for all temperatures where $\kappa$ is known. The electric field's contribution to $\mathbf{s}_3$ is a curve of $E$ and $D$ with $T$ as the parametric variable. The assumptions are that the isothermal system is linear, on an isobar with $\kappa = \kappa(T)$ and $v = v(T)$ only and the electric field's contribution to the entropy is given by expression (25). The specific equations used are:

$$E = \sqrt{\frac{2\mathcal{S}_3}{v\varepsilon_0 \left.\frac{\partial \kappa}{\partial T}\right|_p}} \ . \tag{26}$$

$E$ is a function of $\mathcal{S}_3$ and $T$. $D$ is determined from equation (1) and (26) as

$$D = \varepsilon_0 \kappa \sqrt{\frac{2\mathcal{S}_3}{v\varepsilon_0 \left.\frac{\partial \kappa}{\partial T}\right|_p}} \tag{27}$$

The values of the dielectric constant versus $T$ are based on the equation $\kappa = \kappa(T, p)$ which describes experimental data in the next section with measurements restricted to isobars.

$E$ and $D$ in equations (26) and (27) are functions of $T$ and the selection of $\mathcal{S}_3$. $n$ values were measured in [24] on an isobar and with the aid of (3) the $T$ dependence of $\kappa$ is now investigated.

*Experimental Data for an Optical Glass*

Experimental $n$ data for calcium aluminate glass of composition 5% $SiO_2$, 41.5% $Al_2O_3$, 5% MgO and 48.5% CaO which is listed as NBS F-75 glass were obtained by measuring the index of refraction every 10 C degrees at selected temperatures from -200 C to 600 C. An interference refractometer measured the optical path length differences and thus $n$ to quite high precision of ± $10^{-7}$ or better. Descriptions of optical path length changes in the material were measured with Fizeau-type interference fringes and generated the numerical values for $n$ as a function of $T$ in degrees C. The experimental values of $n$ used with equation (3) thus establish $\kappa$ values. This allows for precise $\kappa$ values versus $T$. Most glass manufacturers [22] also have index values versus temperature often in graphical form.

The refractive index $n = n(T)$ was given as a simple power series at the wave length of 587.6 nm (helium line) at atmospheric pressure. Waxler's and Cleek's power series was used to numerically tabulate their measured $n$ data which implies the measurements have been smoothed. $n$ is converted to $\kappa$ which is then fit using absolute $T$ values in the power series in $T$ shown in equation (28).

$$\kappa = \kappa(T) = \sum_0^3 b_i T^i . \tag{28}$$

The power series for $\kappa$ is a $3^{rd}$ order fit to the data; the same order as the original data.

The expressions for *n* are wavelength dependent.  The frequencies chosen are most often near those used in optical systems.  The times for heat equilibration at optical frequencies are so short that there is not sufficient time for heat to equilibrate.  These optical frequencies in the dielectric media have wavelengths that cover 1,000s of atoms so the local temperature changes as shown in equation (21) can't equilibrate and the processes are adiabatic.  The reported values are representative of the adiabatic index of refraction rather than the isothermal expression used in equation (2).

The data in figure 3 indicates how well the measured dielectric constant is fit by (28) with respect to the data from [24].  Figure 3 (a) is $\kappa = \kappa(T)$ which is used in expression (27) and the derivative of $\kappa$ with respect to $T$ in equations (26) and (27).  Figure 3 (b) is $1/\kappa$ versus $T$ which will be used below.

### D versus E with $\mathit{s}_3$ Constant

The $T$ dependence of $v$, $\kappa$ and $d\kappa/dT$ was used to construct figure 4 (a).  The figure uses $v(T)$ from Appendix A in both equations (26) and (27).  Figure 4 (a) describes the displacement field versus electric field on the constant $\mathit{s}_3$ curve.  The isentropic points are represented by the dotted curve formed by the isothermal linear dielectric constants.  The isothermal dielectric constants are the slopes of straight lines from each dot that passes through $E = 0$, $D = 0$.  In figure 4 (a) and (b) the adiabatic dielectric constant is the slope of the $\mathit{s}_3$ curve using equation (29); the two extreme adiabatic lines of starting and ending temperatures are both aligned with the starting and ending dots seen in the figure.  The adiabatic dielectric constant $\kappa_{\mathit{s}_3,p}$ is the slope from the dotted line:

$$\left.\frac{\partial D}{\partial E}\right|_{\mathit{s}_3,p} = \varepsilon_0 \kappa_{\mathit{s}_3,p} \qquad (29)$$

The $\mathit{s}_3$ curve's slope as seen in figure 4 (a) and (b) is positive, negative and then positive again.  Figure 4 (c) is a schematic showing these changing slopes.  The adiabatic and "isothermal" dielectric constant values are shown in figure (5) versus $T$.

All the dielectric constants which were treated as "isothermals" from the original recorded data are in figure 4 (a), details at high $T$ are in figure 4 (b) with a schematic in 4 (c).  As noted above $\kappa$ values are not "isothermal" but rather adiabatic; the isothermal slopes are taken as straight lines that pass-through $E = 0$, $D = 0$ as in figure 1; the dots in figure 4 (a) and (b) form the isentropic curve.  Every dot shown on the graph is from the original "isothermal" data after application of equations (26) and (27).  The measured "isothermal" dielectric constants are undoubtedly adiabatic as was discussed above.  The original data comes from an energy-per-unit volume system and in the thermodynamic analysis presented in this manuscript they have been treated as "isothermal". The appendix figure C-1 keeps $v$ constant and is for comparison. Dielectric constants for optical materials are generally listed as a function of $T$ but are not isothermal measurements.

The isobaric, adiabatic permittivity is more applicable to optical frequency electric fields at atmospheric pressure where there is insufficient time for the heat to fully equilibrate within the dielectric media over the wavelength distances within the media. This time scale is set by the time between electric field oscillations as shown in figure 2. The full isobaric, adiabatic electric permittivity is:

$$\frac{1}{v}\frac{\partial \underline{D}}{\partial E}\bigg|_{s,p} = \varepsilon_s = \frac{1}{v}\frac{J(\underline{D}, s, p)}{J(E, s, p)} = \varepsilon - \frac{vE^2 T}{C_{p,E}}\left(\frac{\partial \varepsilon}{\partial T}\bigg|_p\right)^2. \quad (30)$$

The Jacobian entries are again evaluated from the rows in Table 1 with $c_1$ eliminated by equation (15). The isentropic, electric permittivity is more applicable at optical frequencies. All the terms on the far right in (30) are positive or zero so the term is either negative or zero. The heat capacity term, $C_{p,E}$, is always considered positive; the heat capacity $C_{p,E}$ is the total heat capacity in equation (9). The energy of the linear system, $ED/2$ on a constant $s_3$ curve, decreases with increasing $T$ since figure 4 (a) has the -200 C point in the upper right and the curve extends towards the lower left at +600 C. The sign on $\varepsilon_s$ depends on both terms on the right side of (30). The electric field's contribution to heat capacity is now described.

The electric field's heat capacity contribution is now found from the entropy, $s_3$. $C_{3,p,E}$ is the heat capacity from the $T$ derivative of equation (25) with $p$ and $E$ held constant.

$$\frac{\partial s_3}{\partial T}\bigg|_{p,E} = \frac{C_{3,p,E}}{T} = v\varepsilon_0 \frac{E^2}{2}\left[\frac{\partial^2 \kappa}{\partial T^2}\bigg|_p + \alpha \frac{\partial \kappa}{\partial T}\bigg|_p\right]. \quad (31)$$

The second derivative of the dielectric constant with respect to $T$ in equation (31) and seen in figure 3 (a) mostly determines the sign of the heat, capacity $C_{3,p,E}$. The 2$^{nd}$ derivative term on the right changes sign as $T$ increases in figure 3 (a) while the second term in equation (31) is the volumetric thermal expansion of the material from figure A-3 times the first derivative of the dielectric constant with respect to $T$. The thermal expansion times the thermal slope of the dielectric constant term are both positive. The heat capacity with constant $E$ is given above while the constant $D$ heat capacity is below. The constant displacement heat capacity $C_{3,p,D}$ from equation (25) written with $D$ from equation (1) is

$$\left.\frac{\partial \mathbf{s}_3}{\partial T}\right|_{p,D} = \frac{C_{3,p,D}}{T} = -v\frac{D^2}{2\varepsilon_0}\left[\left.\frac{\partial^2(\frac{1}{\kappa})}{\partial T^2}\right|_p + \alpha\left.\frac{\partial(\frac{1}{\kappa})}{\partial T}\right|_p\right]. \qquad (32)$$

The reciprocal dielectric constant versus $T$ is in figure 3 (b). Both heat capacities influence changes in the adiabatic dielectric constant curve of figures 4 and 5 although there is some modification by the second terms on the right in expressions (31) and (32). These curves include the $T$ dependence of $v$. Both curvature and thermal expansion contribute to the component of the heat capacity due to the electric field. See additional discussion in the Conclusions.

The total entropy in this system is from Table I by reading the $\mathbf{s}$ row with the independent variables $T$, $p$ and $E$ and then integrating

$$\mathbf{s} = \mathbf{s}_1 + \mathbf{s}_2 + \mathbf{s}_3 = \int_0^T \frac{C_{p,E}}{T}dT - \int_0^p v\alpha\,dp + \int_0^E v\left.\frac{\partial\varepsilon}{\partial T}\right|_p EdE. \qquad (33)$$

The term on the far right was already evaluated in equation (25). The two other parts of the entropy are from the heat capacity integrated over $T$,

$$\mathbf{s}_1 = \int_0^T \frac{C_{p,E}}{T}dT, \qquad (34)$$

and the thermal expansion, $\alpha$, times $v$ integrated over pressure.

$$\mathbf{s}_2 = -\int_0^p v\alpha\,dp. \qquad (35)$$

*Conclusions*

A formal thermodynamic analysis has been developed describing linear dielectric solids with the independent variables $T$, $p$ and $E$. The Gibbs free energy per-unit mass was obtained using a new state variable, the displacement volume, $\underline{D}$. The number of atoms in an energy-per-unit mass system when heating, cooling or pressurizing is constant; the atoms that enter or leave the per-unit volume system each have chemical potentials that contribute to the energy in the system. This energy is not accounted for in volume based systems. The ratio of thermal energy to electrification energy for isothermal linear dielectrics is small at about $10^{-3}$. This implies that all thermal effects are small. The measurements of refractive indexes at optical frequencies as presented here are isentropic yet are reported with no distinction from isothermal values in the literature. Isothermal, dielectric constants require time for heat equilibration. The optical dielectric constants are due to electronic resonances at optical frequencies. If the frequencies were to approach zero for isothermal conditions to prevail then the mechanisms would change and a fully isothermal dielectric constant would have a significantly different dielectric value.

Thus, the isothermal dielectric permittivity values implied in equation (2) don't exist for optical media. Yet $T$ is one of the key considerations for the index of refraction of materials in optical physics.

Isothermal and adiabatic dielectric constants differ as seen by equation (30). The second term on the right in equation (30) was initially thought to be small because thermal effects are small in linear dielectrics. Yet, using the measured $T$ dependence of $\kappa$ as seen in figure 3, the value of $K_{\sigma_3}$ can be constructed from figure 4 (a), (b) and (c). Figure 5 shows large variations in $K_{\sigma_3}$ on the $\sigma_3$ curve for the temperatures investigated. Most of the adiabatic slopes are positive until point marked ② in 4 (c). At this point $K_{\sigma_3}$ from equations (26) and (27) is seen to shows several major irregularities. Figure 4 (b) has the dielectric constant with double values at high temperatures and other oddities: negative heat capacities and negative dielectric constants that can't exist. The relation between heat capacities and the $T$ dependence of $\kappa$ is further developed below.

The heat capacities due to electrification as derived in equations (31) and (32) predict negative values at the temperatures where the second derivative of $\kappa$ and $1/\kappa$ with respect to $T$ change sign. $\kappa$ and $1/\kappa$ in figure 3 (a) and 3 (b) change sign at 722 K and 714 K respectively from the curve fit data. Ignoring for the moment, the second terms on the far right in equations (31) and (32) and using figure 4 (c) for the $\sigma_3$ curve we see at ② the following extremums:

$$\left.\frac{\partial D}{\partial E}\right|_{\sigma_3,\rho} = 0 \; ; \; \left.\frac{\partial D}{\partial T}\right|_{\sigma_3,\rho} = 0 \qquad (36)$$

and at point ③ which is at a higher temperature we have

$$\left.\frac{\partial D}{\partial E}\right|_{\sigma_3,\rho} = \infty \; ; \; \left.\frac{\partial E}{\partial T}\right|_{\sigma_3,\rho} = 0 \qquad (37)$$

The electronic components of the heat capacities from higher $T$ values are negative. The electronic component of the heat capacity from about 714 K to higher $T$ values for constant $D$ field is negative in equation (32); the heat capacity from about 722 K to higher $T$ values for constant $E$ field is negative in equation (31). Above 722 K both heat capacity values are negative. The changes in curvature are barely seen in figures 3 (a) and (b) but easily found from the curve fit data; changes in the sign of these heat capacities correspond to unstable system temperatures. The term $\alpha \left.\frac{\partial \kappa}{\partial T}\right|_P$ is taken to be small; compare figure 4 (b) with figure C-1. In equation (30), if $\varepsilon_0 K_\sigma$ were to be replaced with $\varepsilon_0 K_{\sigma_3}$ then the heat capacity $C_{p,E}$ in the denominator on the lower right is replaced with $C_{3,p,E}$. When the heat capacity $C_{3,p,E}$ becomes zero, the slope on the $s_3$ curve becomes infinite i.e., at point ③. Thus, the change in sign of the 2$^{nd}$ derivative in $\kappa(T)$ dominates the shape of the $s_3$ curve. This is easily seen in figure 5. Also when the heat capacity $C_{3,p,D}$ is zero then the slope on the $s_3$ curve becomes zero i.e., the point ②. Thus, the change in sign of the 2$^{nd}$ derivative in $1/\kappa(T)$ changes the sign on $\varepsilon_0 K_{s_3}$. The electric field's heat capacities change sign depending on the dielectric constant or the reciprocal dielectric constant curvature changing sign. The thermodynamic analysis of the temperature dependence of the isometric dielectric constant for isobaric systems is at odds with physical and thermodynamic laws since the points ② and ③ are unstable thermodynamic points.

The total entropy of a dielectric media is given by equation (33). $s_1$ (T, p, E) and $s_2$ (T, p, E) in the general case are additional entropy terms that will modify the shape and character of figures 4 (a)-(c) since $s_3$ will no longer be constant. An isobaric system at zero pressure eliminates $s_2$ but $s_1$ will still contribute to the entropy. Figure 5's sensitivity to the instabilities driven by the curvature change of the dielectric constant versus $T$ will be eliminated if the heat capacity includes all contributions from equation (8). $s_1$ is a function of $T$ and if included in equation (25) then a constant $s$ line does not exist due to the additional state variables. The isothermal lines can cross at $E = 0$, $D = 0$ as seen in figure 1 but adiabatic lines can't cross without thermodynamic violations. The value of $s_3$ can be made very large with respect to $s_1$ by choosing very large electric fields. Figure 4 can always be considered as a limit with $s_1$ making a minor contribution to the electronic entropy.

The $T$ dependence of the index of refraction has been from the very earliest investigations been considered to be described through the density i.e., the reciprocal specific volume. Lorentz and Lorenz and Drude based expressions [5-10] that connect the dielectric permittivity to the density are noted below in equation (38). In that expression, the index of refraction is often thought to be directly dependent on $T$ and indirectly dependent on $T$ through $v(T)$ i.e., $n(T, v(T))$. This is a thermodynamic relationship which has been explored with numerical values of $n(T, v(T))$ [9, 24]. Some materials have shown that when using this concept both parts make important contributions to $n$. The index of refraction at optical frequencies was thought to be closely connected to $v$ through

$$n = n(T, v) \qquad (38)$$

The numerical values attempted to separate the direct and indirect contributions to $n$. Unfortunately, values were developed without the full thermodynamic investigation as presented here. All numerical values of $n$ at optical frequencies are based upon isobaric measurements while the $T$ dependence of equation (38) mandates that direct measurements should be isometric at optical frequencies with varying *T*. Furthermore, optical measurements are adiabatic not isothermal as equation (38) implies.

A thermodynamic system can be defined with *T*, *v* and *E* as the independent state variables. See equation (5) for constructing a Helmholtz free energy function which can be used to obtain the characteristics of such a new system. In this hypothesized system *v* is an independent variable. $\varepsilon$ is defined on an isometric line, not the isobaric line used in equation (11) in this text. However measured *n* values versus *T* are from isobars. The intersection of an isobaric line with an isometric line is a point in phase space with *T* constant. Thus, the equations describing the index and its temperature components are found inconsistently.

The supposition that dielectric constants only depend on *v* has been raised before and is well included in [9, 24] and many other works. Chapter V in [9] argues that the thermal expansion dominates thermal changes in the refractive index. The astute reader will note that figures 3 (a) and A-1 have very similar shapes. If the *v* superstition were supported then the equation for $\varepsilon$ in equation (2) is not a correct description. It is proposed that (39) is not correct but that the dielectric constants are related to the specific volume with

$$\varepsilon = \varepsilon(v); \quad v = v(T, p) . \tag{39}$$

The electric permittivity being only dependent on *v* is not a thermodynamic relation with *v* being dependent on *T* and *p* as a thermodynamic relation. Expression (39) eliminates all the problems associated with thermodynamic violations arising from equation (25).

The linear dielectric seen in figure 1 has the dielectric constant increasing with increasing *T*. Chatelier's principle [27] states that when a process takes place in a system that disturbs the equilibrium then the reaction is always to lessen the forces and energy in the system as it returns to equilibrium. In all linear elastic materials the elastic stiffnesses soften with increasing temperature. The linear optical dielectrics increase the dielectric constant when the temperature increases in most systems. The dielectric constant in optical materials as seen in figure 1 and equation (2) are not followed in all glasses and are not isothermal but are isentropic so Chatelier's principle is not necessarily comprised nor anomalous. An isothermal system should have the displacement field decreasing as *T* increases so a cycle can't be constructed that generates electrical work by moving heat from a low temperature source to a high temperature source. The dielectric constant versus *T* is self-consistent with the thermodynamics developed here at *T* values less than about 720 K. Dielectric values above 720 K show negative heat capacities; below 720 K Chatelier's principle seems to be anomalous and Kelvin-Planck as explained above and in appendix D is comprised. Negative dielectric constants are seen in figure 4 (c) in the very narrow temperature range of $\pm 10K$ near 720 K the adiabatic dielectric constant taken as the local slope of the $\varepsilon_3$ curve is negative. The physical interpretation of a negative dielectric constant implies electrons attract and move in directions inconsistent with physical laws.

The expressions developed are based on energy-per-unit mass via equation (5) -- the first law of thermodynamics. The introduction of $\underline{D}$ as a state variable through equation (7) is unique and not used by others aside from my work in [21]; it mathematically allows the Gibbs energy function in an energy-per-unit-mass to be found in equation (8). Yet, the definitions in equations (12) through (14) and equation (1) are from energy-per-unit volume concepts. Most of the electro-magnetic community including all Maxwell's laws of electro-magnetism use energy-per-unit volume.

The use of energy-per-unit volume as opposed to energy-per-unit mass unintentionally places the hidden restriction of constant volume on the third order Gibbs functions; the permittivity when constrained to a constant volume system is not the conventional constant pressure thermodynamic systems used in piezo-electric thermodynamics. A simple way to go from a mass to a volume energy system is to hold $v$ constant. If the definitions chosen in Table I are to be consistent with energy per-unit-volume, then $v$ would be constant. In Table I holding $v$ constant mandates the system to be incompressible, the thermal expansion coefficient to be zero and the compressibility of the permittivity to be zero. In this text that application directly effects equations (30) and (31) since $v$ is to be considered as a constant then $\alpha$ should be zero.

Third-order derivatives of the Gibbs function are listed in Appendix B. They were used to establish the thermodynamic identities seen in Table III. Table III when compared to third-order derivatives of the energy-per-unit volume from a Gibbs like function in reference [16, 26] can only be reconciled when $v$ is constant. See Appendix B and Table III.

The second major finding of the paper is that the dielectric constant is dependent only on the volume through equation (39). Furthermore, we have from equations (3), (30) and (39) that

$$\kappa_\sigma = \kappa(1 - \frac{2(\kappa')^2}{\kappa(\kappa'' + \alpha\kappa')}) \qquad (40)$$

Choosing $\alpha$ as zero as noted above and see appendix B plus table III, gives an explicit differential equation for $\kappa$:

$$\frac{2(\kappa')^2}{\kappa(\kappa'')} = \text{constant} \qquad (41)$$

The solution to (41) will have two arbitrary constants and is

$$\frac{\kappa}{\kappa_0} = \frac{v^m}{v_0^m} = \frac{\rho_0^m}{\rho^m} \qquad (42)$$

In equation (42) $\rho$ is the density. The dielectric constant when being described by (42) does not directly depend on $T$ so all the thermodynamic violations noted are eliminated. Figure 6 is a plot of the experimental data for NBS F-75 glass shown as a $ln(\kappa)$ vs $ln(v)$ graph. The slope is $m$.

Appendix D is a selection of additional materials that are in the same format as figure 6. There are two glasses: Phosphate glass F-1329 and Germanate glass F-998. Empirical evidence from all three graphs supports:

$$\frac{\kappa}{\kappa_0} = \sqrt{\frac{v}{v_0}} \qquad (43)$$

Expression (42) implies that the $T$ and $p$ dependence of the dielectric constant is given by

$$\left.\frac{\partial k}{\partial T}\right|_p = \left.\frac{d\kappa}{dv}\frac{\partial v}{\partial T}\right|_p = m\kappa\alpha \qquad (44)$$

and

$$\left.\frac{\partial k}{\partial p}\right|_T = \left.\frac{d\kappa}{dv}\frac{\partial v}{\partial p}\right|_T = -m\kappa\beta \qquad (45)$$

*Acknowledgements*

I'd like to thank Govind Agarwal and Joseph Eberly for reviewing this manuscript; also Robert Boyd and Wayne Knox for encouragement. Furthermore, Ryan Rygg, Gilbert Collins, Danae Polsin, Suxing Hu and colleagues at LLE working on laser-shocked-materials within the High Energy Density Science program. Their efforts are designed to achieve 100 TPa of pressure and identify materials that are only dreamed, thus discovering materials still unknown to mankind. Also, thanks go to Sean Burns at the National Center for Atmospheric Research and John Lambropoulos at the University of Rochester for long term support and encouragement.

Table I. A description of the physical properties of dielectric material with $T$, $p$ and $E$ as independent variables in Jacobian format

| function | $\left.\dfrac{\partial}{\partial T}\right|_{p,E}$ | $\left.\dfrac{\partial}{\partial p}\right|_{T,E}$ | $\left.\dfrac{\partial}{\partial E}\right|_{T,p}$ |
|---|---|---|---|
| g | -s | v | -$\underline{D}$ |
| T | 1 | 0 | 0 |
| p | 0 | 1 | 0 |
| E | 0 | 0 | 1 |
| s | $C_{p,E}/T$ | $-v\alpha$ | $vc_1$ |
| v | $v\alpha$ | $-v\beta$ | $vc_2$ |
| $\underline{D}$ | $vc_1$ | $vc_2$ | $v\varepsilon$ |

Table II. The isothermal ratio of heat to electrical work for several optical glasses as is given by equation (17). Data is from [22] the e-line in Schott's Technical Information, Advanced Optics, TIE-19 July 2016 pages 3 through 5.

| Material | Optical position | Index of refraction | Temperature coefficient of refractive index | Temperature | Ratio: (heat)/(electrical work) |
|---|---|---|---|---|---|
| P-SF68 glass | flint glass | 2.0052 | $24.1 \times 10^{-6}$ /K | 320 K | $7.7 \times 10^{-3}$ |
| Schott N-BK-7® glass | crown glass | 1.5168 | $3.00 \times 10^{-6}$ /K | 320 K | $1.3 \times 10^{-3}$ |
| N-PK51 glass | crown glass | 1.5285 | $-6.70 \times 10^{-6}$ /K | 320 K | $-2.8 \times 10^{-3}$ |

Table III. Thermodynamic identities among the third order Gibbs' derivatives not restricted to linear dielectrics. Derivatives with respect to $T$ are denoted with 0; derivatives with respect to $p$ are denoted as 1; derivatives with respect to $E$ are denoted as 2. See appendix B for examples.

| Property | $\left.\dfrac{\partial}{\partial T}\right|_{p,E}$ | $\left.\dfrac{\partial}{\partial p}\right|_{T,E}$ | $\left.\dfrac{\partial}{\partial E}\right|_{T,p}$ |
|---|---|---|---|
| $C_{p,E}/T$ | $c_{000}$ | $c_{100}$ | $c_{200}$ |
| $v\alpha$ | $-c_{100}$ | $-c_{110}$ | $-c_{210}$ |
| $vc_1$ | $c_{200}$ | $c_{210}$ | $c_{220}$ |
| $-v\beta$ | $c_{110}$ | $c_{111}$ | $c_{211}$ |
| $vc_2$ | $c_{210}$ | $c_{211}$ | $c_{221}$ |
| $v\varepsilon$ | $c_{220}$ | $c_{221}$ | $c_{222}$ |

*Figures:*

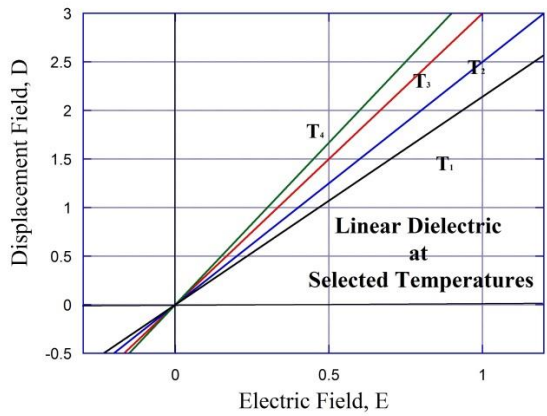

Figure 1. A schematic of a linear dielectric following equation (1) is displayed. The straight lines are isothermal lines. For NBS F-75 glass $T_4>T_3>T_2>T_1$ but this order is reversed for other glasses. The slope changes are exaggerated for this schematic as the slope from $T_1$ to $T_4$ changes by about 1% over 800 C degrees.

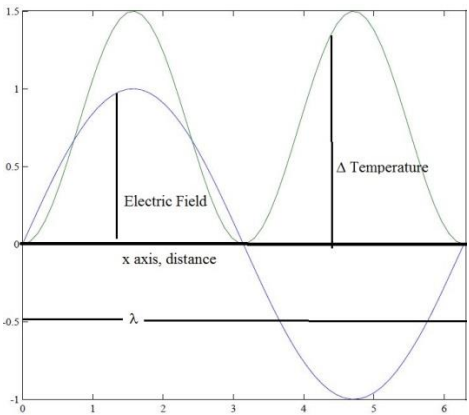

Figure 2. Displaying the temperature change due to an alternating *E* field as predicted by equation (21).

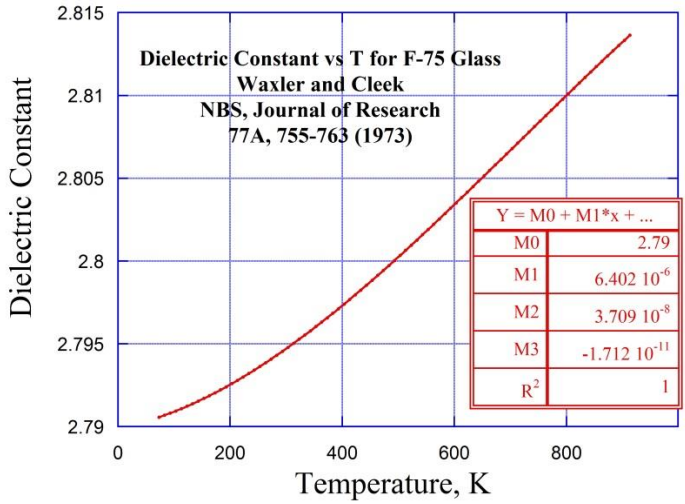

(a)

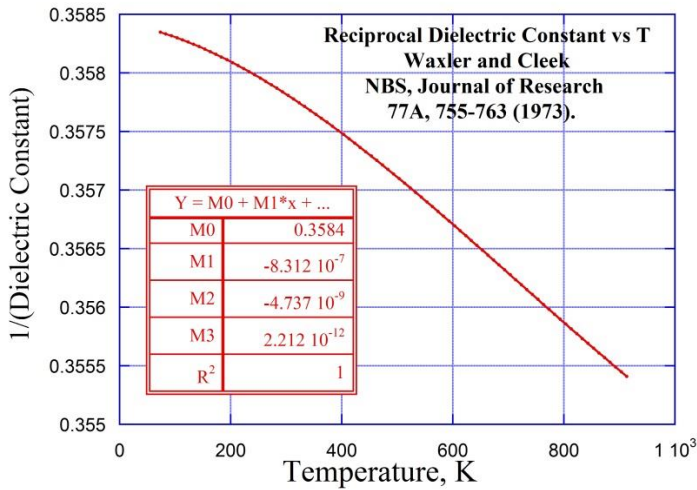

(b)

Figure 3. (a) The temperature dependence of the dielectric constant versus $T$. The original data for the NBS F-75 glass is from [24]. The curve fit data from equation (28) are displayed in each figure. The curvature changes sign at $T$ approximately equals 722 K. (b) Displays $1/\kappa$ or the reciprocal dielectric constant versus $T$. The curvature changes sign when $T$ is approximately equals 714 K. Both $\kappa$ and $1/\kappa$ have changes in the sign of the curvature at high temperatures.

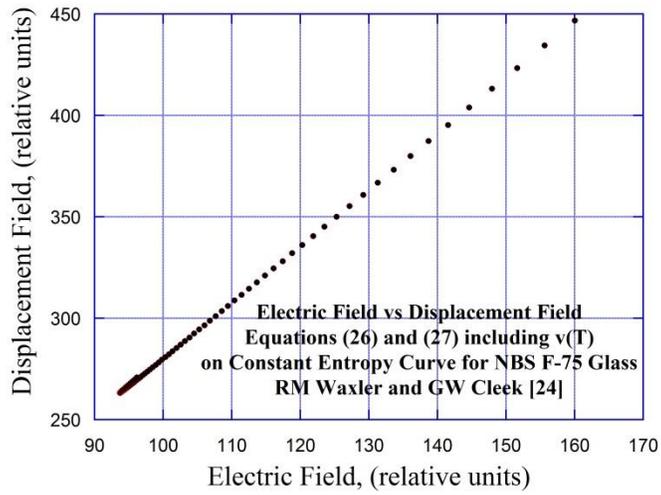

(a)

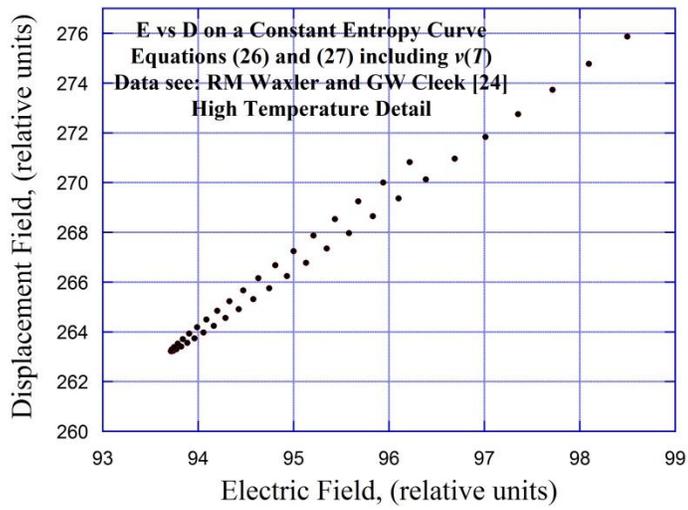

(b)

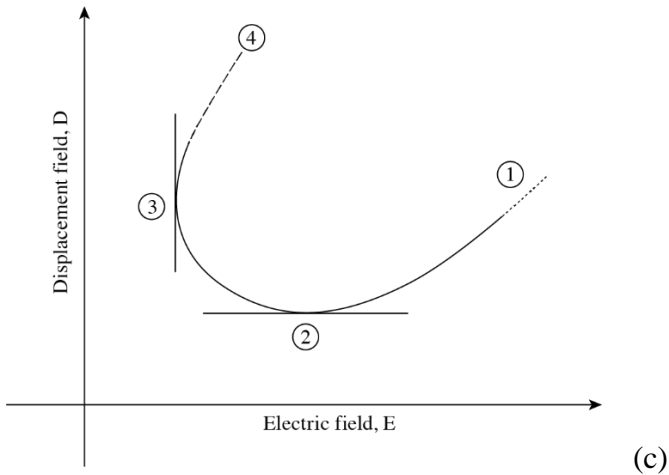

(c)

Figure 4. (a) The constant $\mathcal{b}_3$ curve including $v(T)$ dependence in equation (26) and (27). The units of E are $2\mathcal{b}_3/\varepsilon_0 = 1$ and for D $\varepsilon_0 = 1$. The slope of the curve is the isentropic dielectric constant. See the text for the description of this curve's construction. The low temperature data starting at -200 C is in the upper right; the last point in the lower left is at T = 600 C. (b) Is the lower left part of the curve expanded to show details when the curve changes direction. (c) Schematic of the lower left part of the curve showing a horizontal slope at ② and a vertical slope at ③ on the entropic curve. These two points are explained in the text.

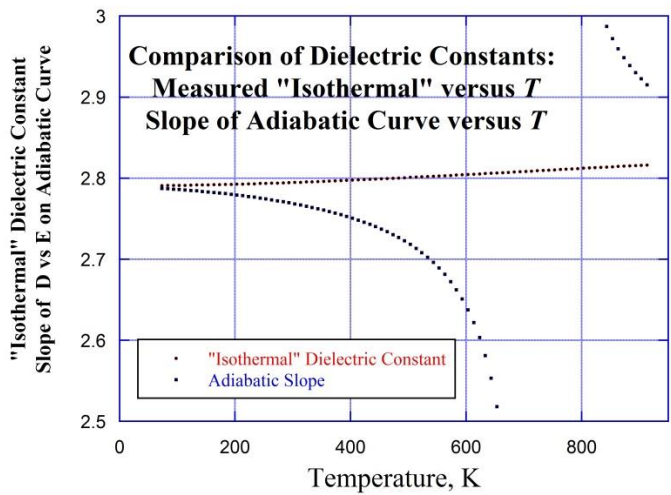

Figure 5. A comparison is shown of the measured "isothermal" dielectric constant as seen in figure 3 (a) and the dielectric constant from the adiabatic slope on the constant $\mathcal{b}_3$ curve. Both curves are displayed versus T.

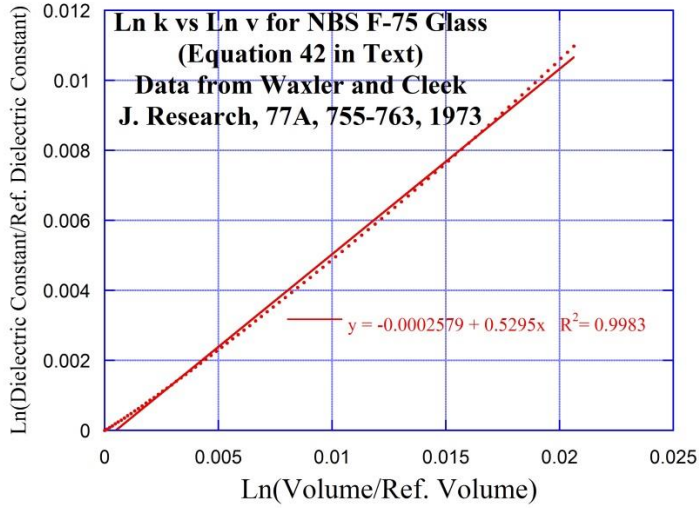

Figure 6. The dielectric constant and volume for NBS F-75 glass are plotted as $\ln(k)$ versus $\ln(v)$. See equation (42) in the text. The original data [24] are parameterized by $T$. The room temperature density of NBS F-75 glass is listed as 2.95 g/cc in [25].

*Appendix A* Expansivity of $v$ was used in expressions (26) and (27). Inclusion of the $T$ dependence of $v$ is from the measured thermal expansivity. Expansivity was used to measure the thermal expansion of the glass NBS F-75. The base measurements were taken directly from [24, 25].

$v(T)$ is found by using the measured linear expansivity given as an engineering strain. The engineering linear strain, $\frac{\Delta \ell}{\ell_0}$, for the glass NBS F-75 is measured versus $T$ in [24]. The volume system is related to the linear strain for isotropic materials by

$$\left(\frac{v}{v_0}\right) = \left(\frac{\ell}{\ell_0}\right)^3. \tag{A-1}$$

The instantaneous length $\ell$ and reference length $\ell_0$ are related by $\ell = \ell_0 + \Delta \ell$. Or we have from equation (A-1):

$$v = v_0 \left(\frac{\ell_0 + \Delta \ell}{\ell_0}\right)^3 = v_0 \left(1 + \frac{\Delta \ell}{\ell_0}\right)^3. \tag{A-2}$$

Figure A-1 (a) is the reported value of $v$ versus $T$ found directly from equation (A-2) and the measured temperature dependence of $\frac{\Delta \ell}{\ell_0}$. Table I uses the volumetric thermal expansion coefficient, $\alpha$. The reference volume is from the reported density of 2.95 g/cc for NBS F-75 glass [25]. The third-order fit to $v$ data plus equation (9) gives the volumetric thermal expansion coefficient seen in figure A-2. $v$ versus $T$ from equation A-2 is used in equations (26) and (27) to create figure 4 (a) and (b).

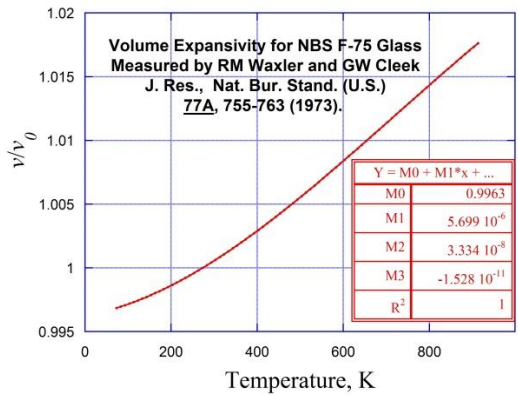

Figure A-1. The volumetric expansivity of NBS F-75 glass is shown versus $T$. $v_0$ was taken as 2.95 g/cc at room temperature [25].

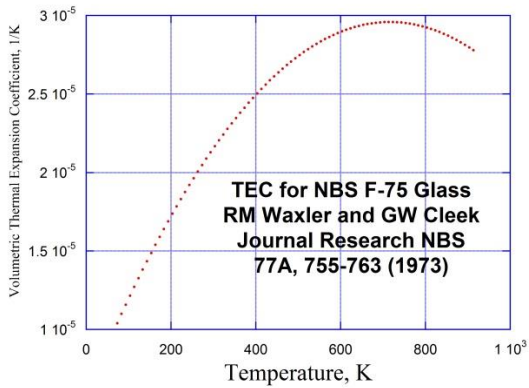

Figure A-2. The volumetric thermal expansion coefficient of NBS F-75 glass is displayed versus $T$.

*Appendix B*  Third-order derivatives of the Gibbs function was used to establish third-order thermodynamic identities and examples are used in Table III. The third-order derivatives of equation (8) are seen in Table III. For example, the three expressions for $c_{210}$ terms when restricted to linear dielectrics are

$$\frac{\partial^3 g}{\partial p \partial T \partial E} = \frac{\partial(v\varepsilon_0 E \frac{\partial \kappa}{\partial T}\big|_p)}{\partial p}\bigg|_{T,E} = \frac{\partial^3 g}{\partial T \partial p \partial E} = \frac{\partial(v\varepsilon_0 E \frac{\partial \kappa}{\partial p}\big|_T)}{\partial T}\bigg|_{p,E} = \frac{\partial^3 g}{\partial E \partial p \partial T} = -\frac{\partial(v\alpha)}{\partial E}\bigg|_{T,p} \quad \text{(B-1)}$$

Properties with duplicate entries are connected as shown in Table III. The terms are of three different types: three derivatives all with the same variable which have no thermodynamic identities; all the mixed derivatives of the variables are shown in equation (B-1); this expression gives two new thermodynamic identities among the three equations that describe the physical properties; two mixed derivatives that give 6 new thermodynamic identities as shown for example for the $c_{200}$ case in equation (B-2) again restricted to linear dielectrics.

$$\frac{\partial^3 g}{\partial E \partial T \partial T} = \frac{\partial(\frac{C_{p,E}}{T})}{\partial E}\bigg|_{T,p} = \frac{\partial^3 g}{\partial T \partial T \partial E} = \frac{\partial(v\varepsilon_0 E \frac{\partial \kappa}{\partial T}\big|_p)}{\partial T}\bigg|_{p,E} \quad \text{(B-2)}$$

In (B-2) the electric field part of the total heat capacity is coupled into the second derivative of the dielectric constant and the thermal expansion coefficient. References [16 and 26] have listed mixed third-order derivatives of piezoelectrics for different crystal symmetry structures. Table III's results will agree with their tables when their crystal system is reduced to an isotropic material and $v$ is held constant; the heat capacity terms that are used in Table III are not included in references [16 and 26]. Equation (B-2) when used in equation (31) relates the total heat capacity to the curvature of the dielectric constant.

$$\frac{1}{T}\frac{\partial C_{p,E}}{\partial E}\bigg|_{T,p} = vE[\frac{\partial^2 \varepsilon}{\partial T^2}\bigg|_p + \alpha \frac{\partial \varepsilon}{\partial T}\bigg|_p] \quad \text{(B-3)}$$

Equations (B-3) and (30) shows that the heat capacity $C_{3,p,E}$ is related to the electric part of the total isobaric, isoelectric, heat capacity. Note $C_{3,p,E}$ is zero when $E$ is zero in both equation (30) and (B-4) which is listed below. The use of energy per unit mass versus energy per unit volume is not just simply multiplying by the specific volume. Models of materials based solely on per unit volume energy systems have been developed and dominate the literature. The differences in third-order derivatives for these two systems are very clear as seen above.

$$C_{3,p,E} = \frac{E}{2} \frac{\partial C_{p,E}}{\partial E}\bigg|_{T,p}. \qquad (B\text{-}4)$$

***Appendix C***  Shown is the parametric *T* dependence of the dielectric constant on the constant $s_3$ curve assuming *v* is a constant.  Exclusion of the *T* dependence of *v* used in expressions (26) and (27) is shown in figure C-1.  Direct comparison of figure 3 (b) to C-1 shows that the far right terms in equations (30) and (31) are small.

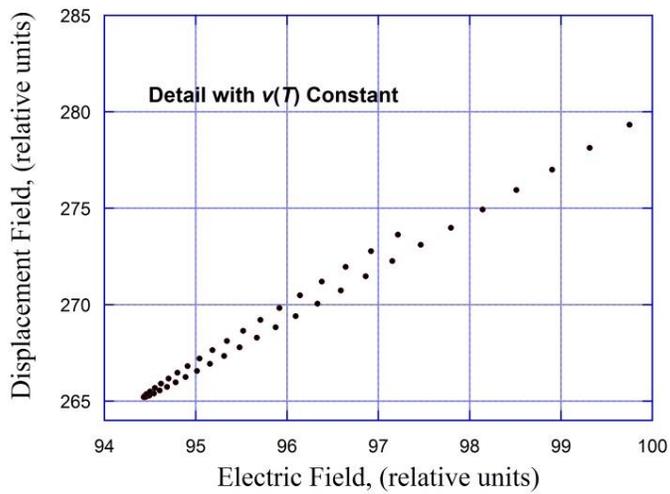

Figure C-1.  The figure is identical to figure 4 (b) except *v* is not kept constant.  The same detailed data but without the *v(T)* dependence from the expansivity directly included in the constant entropy line of $s_3$ again using equations (26) and (27).

***Appendix D***  Additional glasses listed in reference [24] are shown with $ln(\kappa)$ vs $ln(v)$ plotted in format of equation (42).  There are many types of glasses and the thermal/optical behavior of NBS F-75 is considered neither universal nor unique.  Two additional glasses are evaluated and the glasses studied have $m \approx 1/2$.

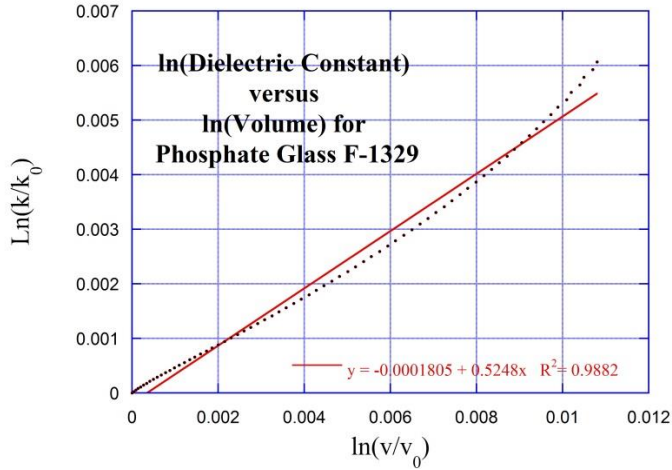

Figure D-1.  $\ell n(\kappa)$ *versus* $\ell n(v)$ for Phosphate glass.

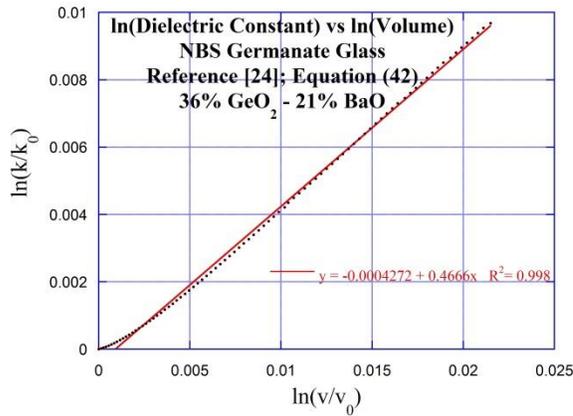

Figure D-2.  $\ell n(\kappa)$ *versus* $\ell n(v)$ for Germanate glass.

*Appendix E*  Several thermodynamic cycles which were developed for the thermodynamic analysis as presented in the text are given in this appendix.  Figure E-1 is a non-linear isothermal curve of *D* versus *E*.  The slopes seen in figure E-1 are the isothermal slope, the isentropic slope, the unstable slope with $\partial D/\partial E\big|_\sigma = 0$ and a second unstable slope with $\partial D/\partial E\big|_\sigma = \infty$. The argument for instability is developed in the text.  Equation (30) relates the slopes of the isothermal and isentropic curves.  The thermodynamic analysis has 8 energy functions with 3 independent variables; instabilities are extremums of these functions.  In figures E-1 to E-3 the ordinate is the state variable, displacement-volume or *v* can be considered constant.

The violation of the Kelvin-Plank second law statement is developed here. If equation 1 were to be replaced with an isentropic electric permittivity as shown in figure E-2 then a second law of thermodynamics violation can be constructed. Optical frequencies are so fast that thermal equilibrium is not achieved over the optical wavelength as seen in figure 2. Figure E-2 is the same as figure 1 except that the optical electric permittivity is assumed now to be isentropic. The expression (E-1) replaces equation (1) as:

$$D = \varepsilon_s E. \qquad (E\text{-}1)$$

Cyclic thermodynamic is used to show thermodynamic violations and that equation (E-1) is unacceptable. Starting with equation (4) for the internal energy $u$, we have for the enthalpy, $h$

$$h \equiv u + pv \qquad (E\text{-}2)$$

so

$$dh = Tds + vdp + vEdD; \quad \oint dh = 0. \qquad (E\text{-}3)$$

The following cycles are all restricted to isobars. Starting with equation (E-1) applied to the figure E-2, the cycle starts at ① and follows an isentropic line to point ②. At point ② the isothermal line which goes from ② to ③ is constructed; see equation (30) and figure E-1. The cycle closes by following the second isentropic line from ③ to ①. The cycle does electrical work and takes heat from the isotherm between ② to ③. This is a violation of the Kelvin-Planck's second law statement. The cycle does electrical work taking heat from a single source. The violation takes place because the two entropy lines cross at $D = 0$; $E = 0$. It follows that equation (E-1) implies a second law violation which is unacceptable.

The second cycle developed is related to the ratio of the electrical work to the heat in a 4 sided cycle. Two isothermal lines and two adiabatic lines define the cycle. It is very similar to the well-known Carnot cycle. The cycle again uses the enthalpy, $h$, seen in equation (E-2) and (E-3). The isobaric cycle is shown in figure E-3 and equation (E-4). The cycle starts at ① goes to point ② on an isotherm. The cycle then follows an isentropic curve from ② to the point ③. At the point ③ the cycle follows the isotherm to the point ④. This closes the cycle while equation (E-5) relates the electrical work to the heat. The clockwise cycle ①->②->③->④ defines a negative electrical work and removes the minus sign on the left in equation (E-4) for the electrical work. The ideal thermodynamic efficiency, $\eta$, of the cycle is a Carnot like efficiency given by equation (E-5) which is a ratio of the electrical work to the heat taken from the high temperature source at $T_2$.

$$-\oint vEdD = \oint Td\sigma = (T_2 - T_1)(\sigma_2 - \sigma_1) \qquad \text{(E-4)}$$

$$\eta = \frac{work}{heat} = (1 - \frac{T_1}{T_2}) \qquad \text{(E-5)}$$

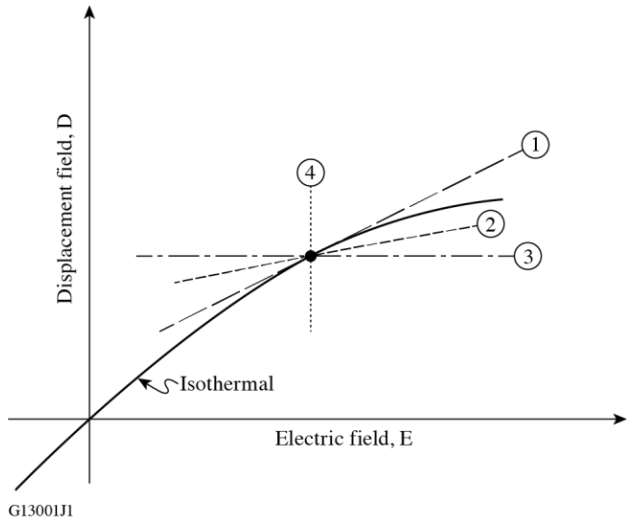

G13001J1

Figure E-1. Non-linear isothermal electrification curve. At a point on the curve: ① is the isothermal slope; ② is the adiabatic slope; ③ is a zero slope and ④ is a vertical slope. See equation (30) for the relations between slopes.

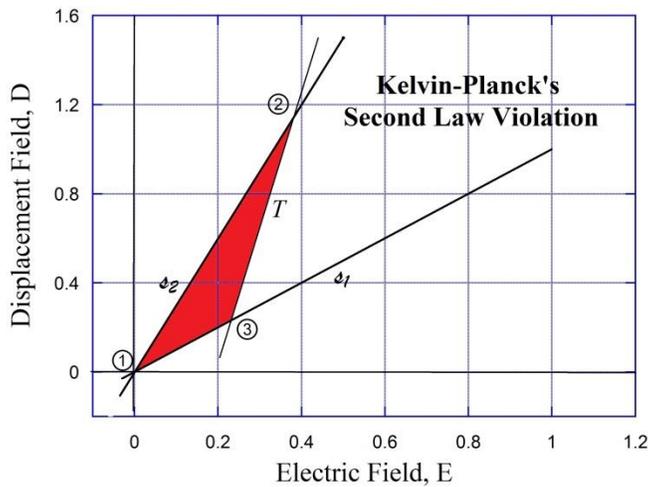

Figure E-2. A schematic cycle that shows two isentropic linear lines that cross at $D = 0$, $E = 0$. The cycle shown violates Kelvin-Planck second law statement as it takes heat from a single source $T$ and converts it into the electric work shown in red.

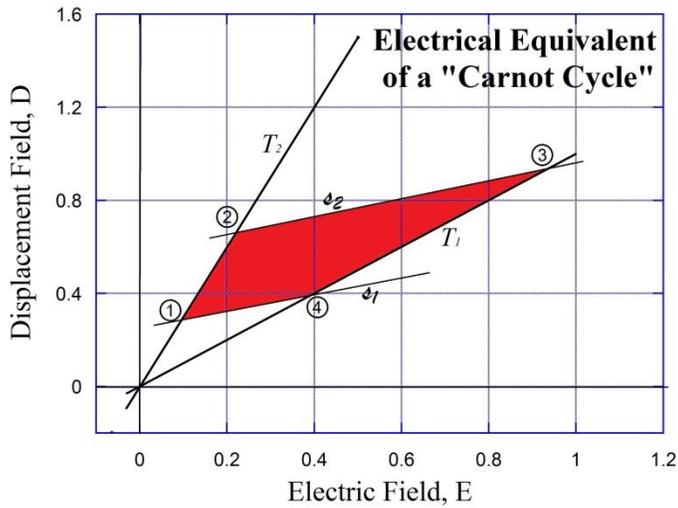

Figure E-3. Carnot like cycle for electrical media showing two isotherms that are connecting two isentropic lines. The electrical work is shown in red. Heat is added and removed only on the isothermal lines.